\documentclass[sigconf]{acmart}

\usepackage{amsmath,amsfonts}
\usepackage{graphicx}
\usepackage{textcomp}
\usepackage{xcolor}
\usepackage{natbib}
\usepackage{hyperref}
\usepackage{multirow}
\usepackage{booktabs}
\usepackage{float}
\usepackage{mathrsfs}
\usepackage{subfigure}
\usepackage{lineno}
\usepackage{algorithm}
\usepackage{algorithmic}
\usepackage{color}

\newtheorem{theorem}{Theorem}[section]

\newtheorem{prop}{Proposition}[section]

\newcounter{hypA}
\newenvironment{hypA}{\refstepcounter{hypA}\begin{itemize}
  \item[({\bf A\arabic{hypA}})]}{\end{itemize}}
\newcounter{hypB}

\newcounter{hypC}

\usepackage{color-edits}
\definecolor{redorange}{RGB}{255, 68, 51}
\addauthor[Denglu]{dl}{redorange}

\AtBeginDocument{%
  \providecommand\BibTeX{{%
    \normalfont B\kern-0.5em{\scshape i\kern-0.25em b}\kern-0.8em\TeX}}}

\setcopyright{acmcopyright}
\copyrightyear{2024}
\acmYear{2024}
\acmDOI{XXXXXXX.XXXXXXX}

\acmConference[KDD'24]{the ACM SIGKDD Conference on Knowledge Discovery and Data Mining 2024}{August 25--29, 2024}{Barcelona, Spain}
\acmPrice{15.00}
\acmISBN{978-1-4503-XXXX-X/18/06}

\acmSubmissionID{806}

\begin{document}

\title{Ego Group Partition: A Novel Framework for Improving Ego Experiments in Social Networks
}


\author{Lu Deng}
\affiliation{%
  \institution{Tencent, Inc.}
  \city{Shenzhen, Guangdong}
  \country{China}}
\email{adamdeng@tencent.com}

\author{JingJing Zhang}
\affiliation{%
  \institution{Tencent, Inc.}
  \city{Shenzhen, Guangdong}
  \country{China}}
\email{broccozhang@tencent.com}

\author{Yong Wang}
\affiliation{%
  \institution{Tencent, Inc.}
  \city{Shenzhen, Guangdong}
  \country{China}}
\email{darwinwang@tencent.com}

\author{Chuan Chen}
\affiliation{%
  \institution{Tencent, Inc.}
  \city{Shenzhen, Guangdong}
  \country{China}}
\email{chuanchen@tencent.com}

\renewcommand{\shortauthors}{Lu Deng, Jingjing Zhang, Yong Wang, and Chuan Chen}

\begin{abstract}
Estimating the average treatment effect in social networks is challenging due to individuals influencing each other. One approach to address interference is ego cluster experiments, where each cluster consists of a central individual (ego) and its peers (alters). Clusters are randomized, and only the effects on egos are measured. In this work, we propose an improved framework for ego cluster experiments called ego group partition (EGP), which directly generates two groups and an ego sub-population instead of ego clusters. Under specific model assumptions, we propose two ego group partition algorithms. Compared to the original ego clustering algorithm, our algorithms produce more egos, yield smaller biases, and support parallel computation. The performance of our algorithms is validated through simulation and real-world case studies.
\end{abstract}

\ccsdesc{Probability and Statistics~Experimental design}
\ccsdesc{Mathematics of computing~Exploratory data analysis}

\keywords{
causal inference, interference, network effects, global average treatment effect, design of experiments, ego experiment
}



\maketitle

\section{Introduction}
Many tech companies today use online controlled experiments to efficiently test and estimate the effects of new product strategies \cite{kohavi1, kohavi2}. Within the Rubin Causal Model framework \cite{rubin}, given certain assumptions, an online controlled experiment can provide unbiased estimates of a strategy's true global average treatment effect (GATE). A critical assumption in the Rubin Causal Model is the Stable Unit Treatment Value Assumption (SUTVA, see \cite{imbens}), which posits that an individual's behavior in the experiment depends solely on their own treatment, not on others' treatments.

Consider a finite population of $n$ individuals.  Let $W_i \in \{0,1\}$ denote the random binary treatment assignment for individual $i$, with 1 and 0 representing treatment and control, respectively. Let $\mathbf{W} = (W_1,\dots,W_n)^{\intercal} \in \Omega \overset{\Delta}{=}  \{ 0, 1 \}^n$ be the experiment assignment vector for all individuals. The potential outcome for individual $i$ when treatment vector $\mathbf{W}$ is implemented is denoted by $Y_i(\mathbf(W))$. The GATE is the difference in population mean when all individuals are treated versus controlled, expressed as
\begin{equation}\label{equa:gate}
\tau \overset{\Delta}{=} \frac{1}{n} \sum_{i=1}^n \Big[Y_i(\mathbf{W} = \mathbf{1}) - Y_i(\mathbf{W} = \mathbf{0})\Big].
\end{equation}
where $\mathbf{1} = (1,\dots,1)^{\intercal}, \mathbf{0} = (0,\dots,0)^{\intercal}$. In what follows, expectations $\mathbb{E}$ are over the treatment assignment only, potential outcomes are considered fixed. Under SUTVA, $Y_i(\mathbf{W}) = Y_i(W_i)$ for all $\mathbf{W} \in \Omega$, simplifying GATE to
\begin{equation}
    \tau = \frac{1}{n} \sum_{i=1}^n \Big[Y_i(W_i = 1) - Y_i(W_i = 0)\Big].    
\end{equation}
A common estimator for $\tau$ is the difference-in-means estimator, defined by
\begin{equation}
\widehat{\tau} = \frac{\sum_{i=1}^n W_i Y_i}{\sum_{i=1}^n W_i}  - \frac{\sum_{i=1}^n (1-W_i) Y_i}{\sum_{i=1}^n (1-W_i)}.
\end{equation}
Under Bernoulli randomization or complete randomization, $\widehat{\tau}$ is an unbiased estimator for $\tau$ \cite{imbens}.

However, in social network settings, the behavior of individuals is likely to be influenced by their social neighbors. For example, if a person frequently uses a new feature, their friends may be affected and also use the feature more frequently. This phenomenon is commonly referred to as network effects, also known as spillover effects or social interference \cite{eckles, abadie2019, kakhki, gui, ugander}. Under such circumstances, $\widehat{\tau}$ is biased and fails to provide a valid estimate of  $\tau$\cite{eckles, gui}. There are usually two paradigms available to address interference: the design-based approach and the model-assisted approach.

In social network settings, individuals' behavior can be influenced by their social neighbors. For instance, if a person frequently uses a new feature, it may prompt their friends to use the feature more often. This is often referred to as network effects, spillover effects, or social interference \cite{eckles, abadie2019, kakhki, gui, ugander}. In such scenarios, $\hat{\tau}$ is biased and does not provide a valid estimate of $\tau$ \cite{eckles, gui}. Two paradigms typically address interference: the design-based approach and the model-assisted approach.

In the design-based approach, the focus is often on the assignment probability  $\mathbb{P}(\mathbf{W}): \Omega \rightarrow [0,1]$. Estimands are typically functions of the potential outcomes $\hat{\tau} = g(Y(\mathbf{W}))$,  with the aim to find a pair $(\hat{\tau}, \mathbb{P}(\mathbf{W}))$ such that the estimator $\hat{\tau}$ has desirable properties under the design $ \mathbb{P}(\mathbf{W}))$ for finite samples. A common method is graph cluster randomization \cite{gui, ugander, Ugander2023, leung2022rate, candogan2021near}, which divides the network graph into clusters and assigns treatment by cluster. The bias of such experiments is linked to the proportion or number of edges between different clusters, making it unfeasible for highly connected networks. However, clustering reduces statistical power, and the bias reduction can be insignificant compared to the increased noise \cite{hayes}. Another prevalent approach is the ego-cluster experiment proposed by LinkedIn \cite{saintjacques}, where a cluster consists of an "ego" (a focal individual) and their "alters" (individuals directly connected to them). The ego clustering algorithm ensures that most of the ego's neighbors are in the same cluster. Randomization is performed on clusters, and inference is made on the ego sub-population. However, the algorithm's inability to generate a large number of egos results in low experimental power.

In the model-assisted approach, potential outcomes are typically modeled as  $Y_i(\mathbf{W}) \sim  F(\theta)$, with estimands being functions of the parameters $\hat{\tau} = g(\theta)$ \cite{ sarndal2003, basse2017, toulis2013, basse2015optimal, sussman2017elements, forastiere2021identification, saint2019method}. A popular model is the exposure model \cite{manski2013, aronow2012}, which maps the full set of treatments to a low-dimensional representation. The elements of this function's codomain, known as "exposures", are used to interpret contrasts among outcomes under different exposures as causal effects. The key assumption for estimation using the exposure model is that the exposures fully describe the causal structure, a condition that may not always be met in reality. Recent studies explore misspecified exposure mappings \cite{chin2018, savje2021average, savje2023causal, leung2022causal}. They examine the standard estimator for $\tau$, which can estimate meaningful exposure effects unbiasedly even without limited interference, indicating a certain robustness to more general interference patterns. However, identifying the conditions for robust inference remains a challenging question.

This paper focuses on the ego-cluster experiment framework, outlining several algorithms to enhance ego network experiments' performance under mild model assumptions. Unlike the original ego clustering algorithm, our approach, called ego group partition (EGP), directly generates the ego sub-population and treatment vector. To ensure the selected egos are homogeneous with the population, our approach first randomly selects a subset of individuals from the population as egos, with the remaining individuals serving as alters. We then randomly assign the selected egos to the treatment and control groups according to a Bernoulli distribution. The remaining nodes are allocated to the treatment and control groups based on specific rules, with the aim of assigning the alter to the group that maximizes the benefit for the ego, using the benefit as a proxy for the exposure condition. Our algorithms offer several advantages, including ensuring the homogeneity of selected egos, allowing for the specification of any number of egos, and supporting implementation through parallel computation, which significantly reduces computation time. The proposed framework are easy to setup and are widely run at Weixin's experiment platform. Our simulation results indicate that the bias of ego experiments conducted with these algorithms is optimal under certain model assumptions. In summary, our contributions are as follows:
\begin{enumerate}
\item We propose an easy-to-implement ego group experiment framework for estimating the global treatment effect. This framework supports parallel computation and yields smaller biases compared to the ego cluster experiment.
\item We implement two algorithms designed to optimize the bias of the estimator under several assumptions.
\item We conduct systematic simulations on a social network and compare our method with several recently proposed methods under a range of settings, providing a credible reference for the effectiveness of our method.
\end{enumerate}

The remainder of this paper is structured as follows. Section \ref{sec2} introduces the preliminaries of network interference and ego-network experiments as proposed by LinkedIn. In Section \ref{sec3}, we detail our proposed ego group partition methods designed to enhance the performance of ego experiments. Section \ref{sec4} includes simulation studies and a real data example that corroborate our theoretical findings. Section \ref{sec5} offers a discussion. Appendix A contains the proofs of some of our theoretical results.

\section{Preliminaries}\label{sec2}

\subsection{Network Interference}
In this work, we propose that the interference can be characterized by an undirected graph $G = (V, E)$, where the node set $V = \{1, 2, \dots, n\}$ symbolizes the individuals, and $E=\{ (i, j)\}$ constitutes the collection of edges that signify interference between individuals.  Each graph is represented by an adjacency matrix. We define $A$ as an $n\times n$ symmetric matrix with entries $(A_{ij})_{1\leq i,j \leq n}$ where $A_{ij} \in \{0, 1\}$, $A_{ii} = 0$, and for $i\neq j$,
\begin{equation}
    A_{ij} \overset{\Delta}{=} \begin{cases}
    0 & \text{no edge between i and j} \\
    1 & \text{undirect edge between i and j}.
    \end{cases}    
\end{equation}
For each individual $i$, we let $N_i \overset{\Delta}{=} \{j: A_{ij} = 1 \}$ represent the set of its neighbors in the graph, and $d_i \overset{\Delta}{=} \sum_{j=1}^n A_{ij} $ denote its degree. Despite considering an unweighted graph in this context, our methodologies can be extended to accommodate weighted graphs.

The potential outcome model $Y_i(\mathbf{W})$ aligns with the model delineated in the introduction. Our objective is to devise an experimental scheme that estimates the GATE, as defined in \eqref{equa:gate}, while minimizing the estimation bias to the greatest extent possible.

\subsection{Ego Cluster Network Experiment}
\begin{figure}[htbp]
\centering
\includegraphics[width=0.46\textwidth]{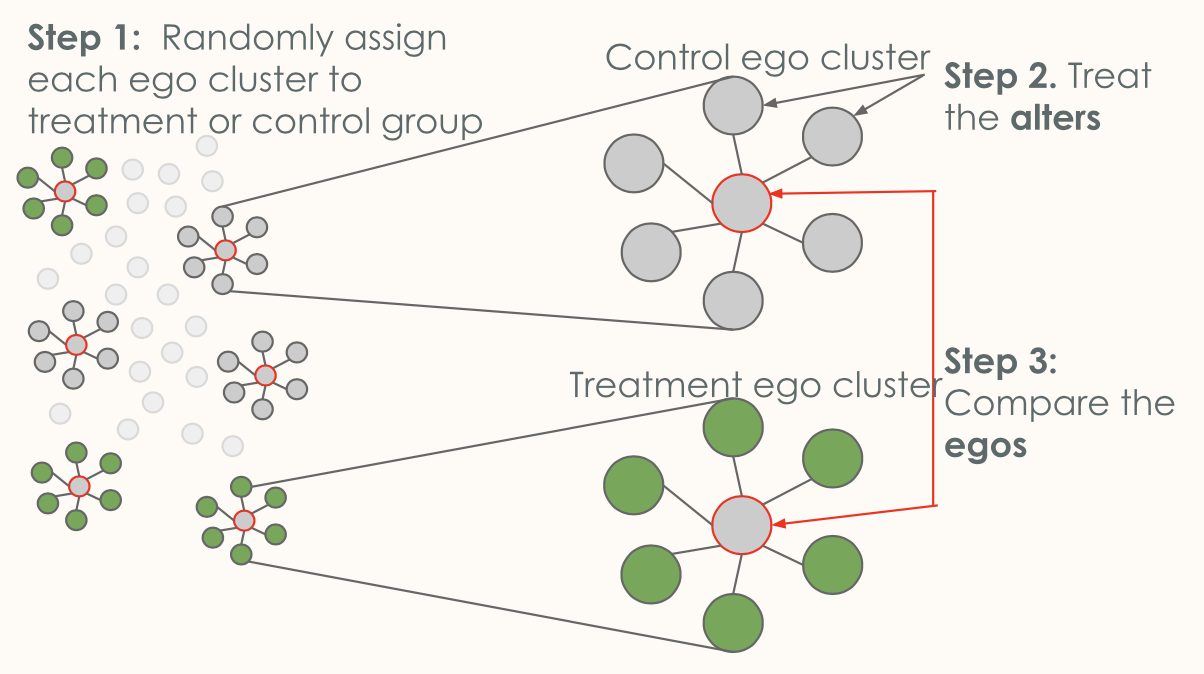}
\caption{An illustration of ego cluster experiment in \cite{saintjacques}.}
\label{egofig}
\end{figure}

LinkedIn \cite{saintjacques} has suggested the application of ego cluster experiments to quantify the spillover effect, circumventing the need for stringent modeling assumptions. In this methodology, the graph is divided into ego clusters, each consisting of a central individual (the ego) and its associated peers (the alters). As illustrated in Figure \ref{egofig}, ego network experiments randomly allocate ego clusters to either the treatment group or the control group. The alters within a cluster are subjected to treatment or control based on the cluster's assigned treatment, while the egos are uniformly assigned to control, treatment, or they share the same assignment as the cluster. During the analysis phase, the comparison is solely between the differences in ego individuals across the two groups, yielding a pure effect estimate.

To elucidate the logic underpinning the treatment allocation in ego cluster experiments, we present the following assumption.
\begin{hypA}
\label{hyp:1}
(Neighborhood Interference) The potential outcomes only depend on one's own and neighbors' treatment assignments: $Y_i (\mathbf{W}) = Y_i (\mathbf{W}^{\prime})$ if $W_i = W_i^{\prime}$ and $W_j = W_j^{\prime}$ for all $j \in N_i$.
\end{hypA}
The neighborhood interference assumption \cite{Ugander2023, liu2022adaptive, yuan2022two, forastiere2022estimating, cortez2022staggered} is frequently employed due to its ability to simplify the problem. The loss rate for each ego is defined as the difference between one and the ratio of the number of alters within the cluster to the ego's total number of peers. The clustering algorithm proposed by LinkedIn is capable of maintaining the loss rate below a predetermined threshold. The objective of ego cluster experiments is to create, for each ego, a network that is treated as though the entire population has undergone treatment.

Nonetheless, ego clustering experiments present several limitations. Firstly, the statistical power of ego network experiments is limited due to the analysis focusing solely on the effects on egos. In practical terms, it is feasible to select merely about 1\% of the individuals as egos. Secondly, the ego clustering algorithm ensures homogeneity between the selected egos and the population by presuming that the outcome is exclusively related to the degree. However, this assumption may not always be valid in real-world scenarios, potentially introducing bias into the results of ego cluster experiments. These constraints have spurred us to devise a more efficient algorithm.

\section{Methodology}\label{sec3}
\begin{figure}[htbp]
\centering
\includegraphics[width=0.4\textwidth]{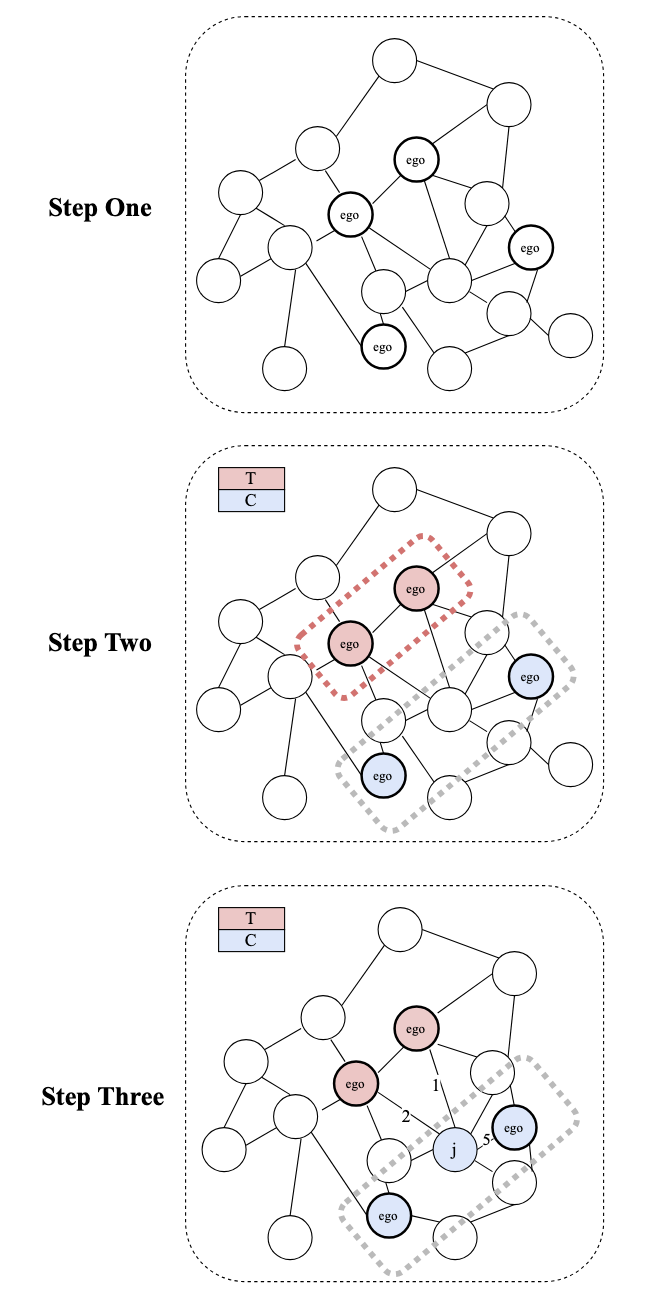}
\caption{Diagram of ego group partition framework.}
\label{ego_framework}
\end{figure}

We initially provide an overview of our proposed ego group partition method (refer to Algorithm \ref{alg: framework}). At the onset of the procedure, a subset of individuals is randomly selected from the population to serve as egos, ensuring homogeneity between the selected egos and the overall population. These selected egos are subsequently randomly allocated to either the treatment group or the control group, following a Bernoulli distribution with a success probability of 0.5. The remaining nodes are then assigned to the treatment group and control group based on predefined rules. During the analysis phase, only ego data is employed for statistical inference. A visual representation of the ego group partition framework is provided in Figure \ref{ego_framework}.

\begin{algorithm}[!ht]
\caption{Framework of ego group partition.}
\label{alg: framework}
\begin{algorithmic}[1]
 \REQUIRE $q\in (0,1)$.
 \ENSURE Ego nodes $E_i$ and treatment assignment $W_i$.
\STATE Randomly select $q$ percent of individuals to be ego, with the remainder alter, without regard to the network structure.
\STATE Randomly assign half of the ego to the treatment group, and the left half to the control group.
\STATE For each of alter $j$, assign alter $j$ to treatment or control group based on specific rules.
\end{algorithmic}
\end{algorithm}

Define $E_i=1$ if individual $i$ is an ego, and $E_i=0$ otherwise. The estimator we utilize is the difference in the average outcome of egos in the treatment and control groups, as given by:
\begin{equation}\label{equa:egoest}
    \widehat{\tau}_{\text{ego}} \overset{\Delta}{=} \frac{\sum_{i=1}^n E_i W_i Y_i}{\sum_{i=1}^n E_i W_i} - \frac{\sum_{i=1}^n E_i (1-W_i) Y_i}{\sum_{i=1}^n E_i (1-W_i)}.
\end{equation}
The variance of this estimator is dictated by the sample size, specifically, the number of egos, which is established in the initial step of Algorithm \ref{alg: framework}. Consequently, to minimize the mean squared error (MSE) of the estimator, our focus should be on minimizing the absolute bias of the estimator. More precisely, we should seek the solution to the following optimization problem:
\begin{equation*}
    \min_{\{W_j: E_j = 0\}} \mathbb{E} \big(\widehat{\tau}_{\text{ego}} - \tau \big) 
\end{equation*}
In light of Assumption \ref{hyp:1}, it becomes necessary to assign the ego's neighboring nodes to the same group as the ego whenever feasible. This requirement calls for some mild assumptions about the potential outcomes. We will proceed to detail the allocation process for the ego's neighboring nodes.

\subsection{Baseline approach}
We start our discussion with a simple linear model, specifically the linear-in-means model without an endogenous effect, as proposed by Cai \cite{cai}. A simplified representation of this model can be expressed as:
\begin{equation}\label{equa:linearmodel}
    Y_i(\textbf{W}) = \beta_0 + \beta_1 \cdot W_i + \beta_2 \cdot \sigma_i.
\end{equation}
In this equation, $\sigma_i =  \sum_{j=1}^n \frac{A_{ij}}{d_i} W_j$ denotes the ratio of treated neighbors for individual $i$. The coefficients $\beta_1$ and $\beta_2$ capture the treatment effect and the spillover effect, respectively. This model is a streamlined version of the more complex model presented in \cite{yu2022estimating}, which accommodates heterogeneous interference effects. Under the framework of this model, we can articulate the GATE in terms of the potential outcome model parameters:
\begin{equation}
    \tau = \beta_1 + \beta_2.    
\end{equation}

Let's consider a scenario where we have already selected a subset of individuals, referred to as the ego subset, and administered a specific treatment to all individuals within this subset. Let $n_1$ and $n_0$ represent the number of egos in the treatment and control groups, respectively. We can then establish the following result:
\begin{prop}\label{prop: basemodelbias}
Given the potential outcome model \eqref{equa:linearmodel}, the bias of ego estimator $\widehat{\tau}_{\text{ego}}$ is
\begin{equation}
   \mathbb{E}(\widehat{\tau}_{\text{ego}}) - \tau = \beta_2 \bigg( \mathbb{E}
   \Big(\frac{1}{n_1} \sum_{i=1}^n E_i W_i \sigma_i - \frac{1}{n_0} \sum_{i=1}^n E_i (1-W_i) \sigma_i\Big) - 1 \bigg).
\end{equation}
\end{prop}

Next, let's define $R \overset{\Delta}{=} \frac{1}{n_1} \sum_{i=1}^n E_i W_i \sigma_i - \frac{1}{n_0} \sum_{i=1}^n E_i (1-W_i) \sigma_i$. We can then propose the following for $R$.

\begin{prop}\label{prop: r_ineuqa}
$ -1 \leq R \leq 1$, and hence $-2 \leq R-1 \leq 0$ .
\end{prop}

With the established proposition, we can infer that to minimize the absolute bias of $\widehat{\tau}_{\text{ego}}$, we should allocate the treatment to the ego's neighbor nodes in a way that maximizes the expectation of $R$. It's important to note that when we proceed with the third step in Algorithm \ref{alg: framework}, the treatment assignment of egos, determined in the second step, remains fixed. Consequently, we propose the following optimization problem:

\begin{equation}
    \max\limits_{ \{W_j : E_j = 0\} } \mathbb{E}\big[R \,\,|\,\, \{W_i: E_i=1 \}\big].
\end{equation}
This optimization problem has a unique solution, which we summarize in the following theorem:

\begin{theorem}\label{thm: r_opt}
The expectation $\mathbb{E}\big[R \,\,|\,\, \{W_i: E_i=1 \}\big]$ achieves its maximum value when all ego's neighbor nodes' treatment assignments are $W_j = \mathcal{I}\{\Delta_j > 0\}$, where 
\begin{equation}
    \Delta_j = \sum_{i=1}^n \frac{ E_i W_i A_{ij} }{ n_1 d_i } - \sum_{i=1}^n  \frac{ E_i (1-W_i) A_{ij} }{n_0 d_i },
\end{equation}
where $\mathcal{I}\{\}$ is the indicator function.
\end{theorem}

In summary, to minimize the absolute bias of the ego estimator $\widehat{\tau}_{\text{ego}}$, we should assign a treatment $W_j = 1$ when $\Delta_j > 0$, and a control $W_j = 0$ otherwise. The specifics of this process are outlined in Algorithm \ref{alg1}. It's worth noting that the third step of the algorithm is particularly well-suited for parallel computation. This means that multiple calculations can be performed simultaneously, significantly reducing the overall running time of the algorithm.

\begin{algorithm}[!ht]
	\caption{Ego group partition with linear model.}
	\label{alg1}
	\begin{algorithmic}[1]
     \REQUIRE $q\in (0,1)$.
     \ENSURE Ego nodes $E_i$ and treatment assignment $W_i$.
    \STATE Randomly select $q$ percent of individuals to be ego, with the remainder alter, without regard to the network structure.
    \STATE Randomly assign half of the ego to the treatment group, and the left half to the control group.
    \STATE For each of alter $j$,
    \begin{itemize}
    \item calculate
        $$
        \Delta_j = \sum_{i=1}^n \frac{ E_i W_i A_{ij} }{ n_1 d_i } - \sum_{i=1}^n  \frac{ E_i (1-W_i) A_{ij} }{n_0 d_i } .
        $$
    \item if $\Delta_j > 0$, assign alter $j$ to treatment group, otherwise to control group.
    \end{itemize}
	\end{algorithmic}
\end{algorithm}

At this point, it's important to highlight a few key aspects of our discussion. Firstly, the term $\sum_{i=1}^n \frac{ E_i W_i A_{ij} }{ n_1 d_i }$ can be interpreted as the relationship between the alter $j$ and the ego treatment group, while $\sum_{i=1}^n  \frac{ E_i (1-W_i) A_{ij} }{n_0 d_i }$ represents the relationship between the alter $j$ and the ego control group. If the relationship between the alter and the ego treatment group is stronger, the alter is assigned to the treatment group, and vice versa. Secondly, the proportion of selected egos has a direct impact on the treated neighbor ratios in each group. This is because the difference in treated neighbor ratios in the treatment and control groups directly influences the bias of the ego experiment, even when the linear potential outcome model \eqref{equa:linearmodel} is misspecified. Selecting a larger proportion of egos results in greater hypothesis-testing power, but a smaller difference in the treated neighbor ratios, and consequently, a larger bias. Therefore, a balance must be struck between power and bias. Lastly, the density of the graph affects the minimum absolute bias of $\widehat{\tau}_{\text{ego}}$. When the proportion of alters with at least two or more ego neighbors is high, it becomes challenging to ensure that alters of egos in two groups are in the same group as the ego. We will propose a method to address such issues in Section \ref{subsec: clustering}.

\subsection{High-level approach}\label{subsec: high-level}
The primary concept underpinning the baseline ego group partition is to minimize the bias of the ego estimator $\widehat{\tau}_{\text{ego}}$ under a linear potential outcome model \eqref{equa:linearmodel}. However, this linear model may be a strong assumption and may not always hold true in reality. A more comprehensive approach involves exposure mapping \cite{manski2013, aronow2012}, which is a function that maps an assignment vector and unit-specific traits to an exposure value: $f:\Omega \rightarrow \Delta$. This function satisfies the following condition:
\begin{equation*}
    Y_i (\mathbf{W}) = Y_i (\mathbf{W}^{\prime}), \text{ if } f(\mathbf{W}) = f(\mathbf{W}^{\prime}),
\end{equation*}
for any $\mathbf{W}, \mathbf{W}^{\prime} \in \Omega$ and any $i$. The codomain $\Delta$ encompasses all possible treatment-induced exposures that might occur in the experiment. Each of the distinct exposures in $\Delta$ may lead to unique potential outcomes for each unit.

To illustrate this with a more tangible example, let's consider the linear model given by \eqref{equa:linearmodel}. In this case, the exposure mapping is defined as $f(\mathbf{W}) = (W_i, \sigma_i)$, and the potential outcome is a linear combination of the two elements in the exposure label. An exposure mapping that allows for completely arbitrary interference would be one for which $\Delta = \Omega$. In this scenario, each unit has a unique type of exposure under each treatment assignment. However, if such an exposure mapping were valid, it becomes evident that estimating the GATE in a meaningful way would be impossible. Therefore, we consider the commonly used exposure mapping $f(\mathbf{W}) = (W_i, \sigma_i)$ as in \eqref{equa:linearmodel}, while allowing the potential outcome to be related to exposure in a more general function. Specifically, we assume the potential outcome is the summation of the functions of $W_i$ and $\sigma_i$:
\begin{equation}\label{equa:convexfunc}
    Y_i(\mathbf{W}) = g_1(W_i) + g_2(\sigma_i), \quad W_i \in \{0,1\}, \,\, \sigma_i \in [0,1].
\end{equation}
Interference is characterized by the properties of the function $g_2$. It is commonly assumed that the function $g_2(\cdot): [0,1] \rightarrow \mathbb{R}$ is monotonic. Without loss of generality, we assume that $g_2$ is non-decreasing. Additionally, for various metrics we test, we find that $g_2$ is a convex function. This means that the function $g_2(\sigma_i)$ usually changes more rapidly when $\sigma_i$ is small and changes more slowly as $\sigma_i$ approaches 1. Formally, we make the following assumption:
\begin{hypA}
\label{hyp:2}
The function $g_2(\cdot): [0,1] \rightarrow \mathbb{R}$ is non-decreasing as well as convex. Its first and second derivatives exist, satisfying $g^{\prime} \geq 0, g^{\prime\prime} \leq 0$.
\end{hypA}

Under the assumption that the function $g_2$ is non-decreasing and convex, we can further optimize the ego group partition algorithm to improve the experimental outcomes. While a rigorous derivation as in the previous section may not be feasible, we can still employ heuristic inferences. Consider the GATE in \eqref{equa:convexfunc}, which is given by:
\begin{equation}
    \tau = \big(g_1(1) - g_1(0)\big) + \big(g_2(1) - g_2(0)\big).
\end{equation}
We propose the following:
\begin{prop}\label{prop: convexmodelbias}
Given the potential outcome model \eqref{equa:convexfunc} and assuming (A\ref{hyp:2}), the bias of the ego estimator $\widehat{\tau}_{\text{ego}}$ can be approximated by:
\begin{align}\label{equa:approxbias}
    \mathbb{E}(\widehat{\tau}_{\text{ego}}) - \tau  & \approx \big(g_2(\mathbb{E}(\sigma_i|W_i=1)) - g_2(\mathbb{E}(\sigma_i|W_i=0)) \big) \nonumber \\
    & \quad \,\, - \big(g_2(1) - g_2(0) \big).
\end{align}
\end{prop}

\begin{figure}[htbp]
\centering
\includegraphics[width=0.4\textwidth]{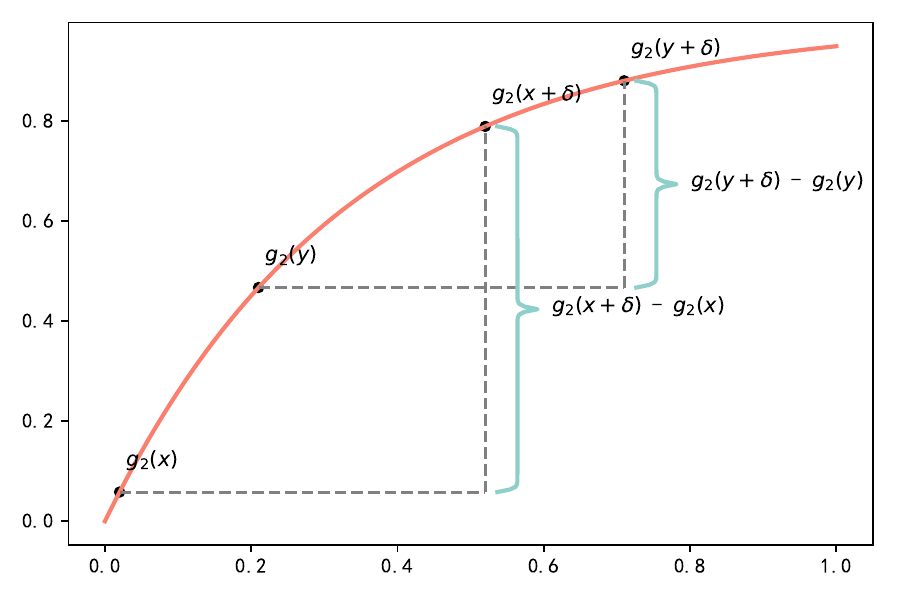}
\caption{An example of convex function $g_2$. For any $0 < x < y < 1$ and $0 < \delta < 1-y$, $g_2(x+\delta) - g_2(x) \geq g_1(x+\delta) - g_1(x)$.
}
\label{exposurefig}
\end{figure}

Given that $g_2$ is non-decreasing, the right-hand side of \eqref{equa:approxbias} is always less than 0. Therefore, to minimize the bias of the ego experiment towards 0, we aim to maximize:
\begin{align}\label{equa: convmaxobj}
   & g_2(\mathbb{E}(\sigma_i|W_i=1)) - g_2(\mathbb{E}(\sigma_i|W_i=0))  \\
   = \,\, & g_2^{\prime}(\xi)\big(\mathbb{E}(\sigma_i|W_i=1) - \mathbb{E}(\sigma_i|W_i=0)\big). \nonumber
\end{align}
Here, $\xi$ lies between $\mathbb{E}(\sigma_i|W_i=0)$ and $\mathbb{E}(\sigma_i|W_i=1)$. According to Assumption \ref{hyp:2}, $g_2^{\prime}(\xi) > 0$. Therefore, we need to make $\mathbb{E}(\sigma_i|W_i=1) - \mathbb{E}(\sigma_i|W_i=0)$ as large as possible, while keeping $\xi$ as close to 0 as possible. When $\mathbb{E}(\sigma_i|W_i=1) - \mathbb{E}(\sigma_i|W_i=0)$ is fixed, the closer $\mathbb{E}(\sigma_i|W_i=0)$ is to 0, the smaller $\xi$ will be. Consequently, the larger $g_2(\mathbb{E}(\sigma_i|W_i=1)) - g_2(\mathbb{E}(\sigma_i|W_i=0))$ will be. This phenomenon is illustrated in Figure \ref{exposurefig}. The figure shows how the difference between the expected values of $\sigma_i$  affects the bias of the ego estimator. As the difference increases, the bias decreases, leading to a more accurate estimation of the GATE. 

Our goal, therefore, is to maximize $\mathbb{E}(\sigma_i|W_i=1) - \mathbb{E}(\sigma_i|W_i=0)$ while keeping $\mathbb{E}(\sigma_i|W_i=0)$ as close to 0 as possible. This would minimize the bias of the ego estimator, leading to a more accurate estimation of the GATE. To achieve this, we propose the following optimization problem:
\begin{equation}\label{equa:max1}
    \max\limits_{ \{W_j : E_j = 0 \}}  \mathbb{E}(\sigma_i|W_i=1) - (1 + \theta) \mathbb{E}(\sigma_i|W_i=0).
\end{equation}
Here, $\theta \geq 0$ is a tuning parameter that prevents $\mathbb{E}(\sigma_i|W_i=0)$ from being too large. This optimization problem has a unique solution, which we summarize in the following theorem. The details of the algorithm are provided in Algorithm \ref{alg2}.

\begin{theorem}\label{thm: conv_opt}
Assume (A\ref{hyp:2}), the optimization solution of \eqref{equa:max1} is $W_j = \mathcal{I}\{\tilde{\Delta}_{j} > \theta\}$, where 
\begin{equation}
    \tilde{\Delta}_{j} = \frac{ \sum_{i=1}^n \frac{ E_i W_i A_{ij} }{ n_1 d_i } - \sum_{i=1}^n \frac{ E_i (1-W_i) A_{ij} }{n_0 d_i } } { \sum_{i=1}^n \frac{ E_i (1-W_i) A_{ij} }{n_0 d_i } }.
\end{equation}
In other words, when $\tilde{\Delta}_{j} > \theta$, we set $W_j = 1$; otherwise, we set it to 0. 
\end{theorem}

\begin{algorithm}[!ht]
	\caption{Ego group partition with convex model.}
	\label{alg2}
	\begin{algorithmic}[1]
     \REQUIRE $q\in (0,1), \theta \geq 0$.
     \ENSURE Ego nodes $E_i$ and treatment assignment $W_i$.
    \STATE Randomly select $q$ percent of individuals to be ego, with the remainder alter, without regard to the network structure.
    \STATE Randomly assign half of the ego to the treatment group, and the left half to the control group.
    \STATE For each of alter $j$,
    \begin{itemize}
    \item calculate
        $$
        \tilde{\Delta}_{j} = \frac{ \sum_{i=1}^n \frac{ E_i W_i A_{ij} }{ n_1 d_i } - \sum_{i=1}^n \frac{ E_i (1-W_i) A_{ij} }{n_0 d_i } } { \sum_{i=1}^n \frac{ E_i (1-W_i) A_{ij} }{n_0 d_i } } .
        $$
    \item if $\tilde{\Delta}_{j} > \theta$, assign alter $j$ to treatment group, otherwise to control group.
    \end{itemize}
	\end{algorithmic}  
\end{algorithm}

To conclude this subsection, we highlight several key points. First, similar to the previous subsection, $\tilde{\Delta}_j$ can be interpreted as the relative difference between the alter's relationship with the ego treatment group and the ego control group. If this relative difference exceeds a certain threshold, the alter is assigned to the treatment group; otherwise, it is assigned to the control group. If $\theta = 0$, this degenerates to the baseline Algorithm \ref{alg2}. Second, as $\theta$ increases, the algorithm is more likely to assign the alter to the control group. Ideally, $\theta$ should be chosen to maximize \eqref{equa: convmaxobj}. However, since it is impossible to know the exact expression of $g_2$ in reality, one can choose an appropriate $\theta$ based on their own real data. Lastly, consider the potential outcome model:
\begin{equation*}
    Y_i = \beta_1 + \beta_2 W_i + \beta_3 \Big(\sigma_i \mathcal{I}\{\sigma_i < \theta\} + \theta \mathcal{I}\{\sigma_i \geq \theta\} \Big),
\end{equation*}
It can be shown that the optimal choice of $\{W_j | E_j = 0\}$  that minimizes the bias is exactly the solution of \eqref{equa:max1}. We leave the proof of this statement to the reader.

\subsection{Second neighborhood clustering approach}\label{subsec: clustering}
In the absence of a convex property imposed on $f(W_i, \sigma_i)$, the primary consideration is to ensure that $\mathbb{E}(\sigma_i|W_i=0)$ is as close to 0 as feasible, while $\mathbb{E}(\sigma_i|W_i=1)$ is as close to 1 as feasible. If the network graph $G$ is not exceedingly sparse, which is the case for most real-world networks, it becomes impossible to fulfill both objectives simultaneously, necessitating some trade-offs between the two goals. In such circumstances, we propose a method that can partially achieve a smaller bias in comparison to the baseline Algorithm \ref{alg1}.

We define $H$ as the matrix that signifies neighbors of neighbors, also known as second neighborhoods, such that
\begin{equation*}
    H_{ij} = \begin{cases}
    1 & \text{if } (i \neq j) \land (A_{ij}=0) \land \big(\sum_{k=1}^n A_{ik} \cdot A_{jk}>0\big) \\
    0 & \text{otherwise}.
    \end{cases}
\end{equation*}
We denote the graph with adjacency matrix $H$ as $G^*$, and the sub-graph of $G^*$ consisting solely of egos as $G^*_{\text{ego}}$. The complexity of optimizing both objectives concurrently, as previously discussed, arises from the edges between the egos of the treatment group and the control group within the ego sub-graph $G^*_{\text{ego}}$. This prompts us to minimize the edges between the egos of the two groups when allocating treatments to egos. Our strategy involves initially partitioning the egos into clusters using appropriate clustering algorithms post ego selection. Following this, we utilize cluster randomization to assign treatment to the egos. We then apply the steps outlined in Algorithm \ref{alg1} to assign treatment to the remaining alters. The specifics of the algorithm are detailed in Algorithm \ref{alg3}.

\begin{algorithm}[!ht]
	\caption{Ego group partition with second neighborhood clustering.}
	\label{alg3}
	\begin{algorithmic}[1]
     \REQUIRE $q\in (0,1)$.
     \ENSURE Ego nodes $E_i$ and treatment assignment $W_i$.
    \STATE Randomly select $q$ percent of individuals to be ego, with the remainder alter, without regard to the network structure.
    \STATE Use a clustering algorithm to partition the second neighborhood graph of egos into clusters.
    \STATE Randomly assign half of the ego cluster to the treatment group, and the left half to the control group.
    \STATE For each of alter $j$,
    \begin{itemize}
    \item calculate
        $$
        \Delta_j = \sum_{i=1}^n \frac{ E_i W_i A_{ij} }{ n_1 d_i } - \sum_{i=1}^n  \frac{ E_i (1-W_i) A_{ij} }{n_0 d_i } .
        $$
    \item if $\Delta_j > 0$, assign alter $j$ to treatment group, otherwise to control group.
    \end{itemize}
	\end{algorithmic}  
\end{algorithm}

While this approach mitigates the bias in ego experiments, it may concurrently amplify the variance of the statistics \eqref{equa:egoest}, leading to a decrease in statistical power. This is attributed to the fact that randomization is executed on ego clusters. Consequently, it is advisable to employ Algorithm \ref{alg2} in practical applications.

\section{Numerical Experiments}\label{sec4}

\subsection{Simulated data}
We carried out a series of simulation studies to juxtapose our proposed algorithms with existing methodologies, specifically for estimation problems occurring in the context of interference. These simulations were executed using a real-world network graph, namely FB-Cornell5 \footnote{Network topology data can be found in https://networkrepository.com/socfb-Cornell5.php }. The FB-Cornell5 network consists of 18,660 nodes and 790,777 edges, resulting in an average degree of 84.7. We implemented three variations of the ego group partition Algorithm \ref{alg2}, each with different thresholds $\theta = 0, 0.2, 0.5$. Alongside our ego group partition approach, we also incorporated the ego cluster approach as proposed by LinkedIn \cite{saintjacques}. To maintain comparable statistical power across different algorithms, we controlled the sample size of egos as the same across all approaches. We selected approximately 2.5\% of individuals in this network to serve as egos. In the ensuing results, the term 'Ego Cluster' denotes the ego cluster approach as proposed by LinkedIn, while 'EGP' signifies our ego group partition approach.

\begin{figure*}[htbp]
\centering
\includegraphics[width=0.7\textwidth]{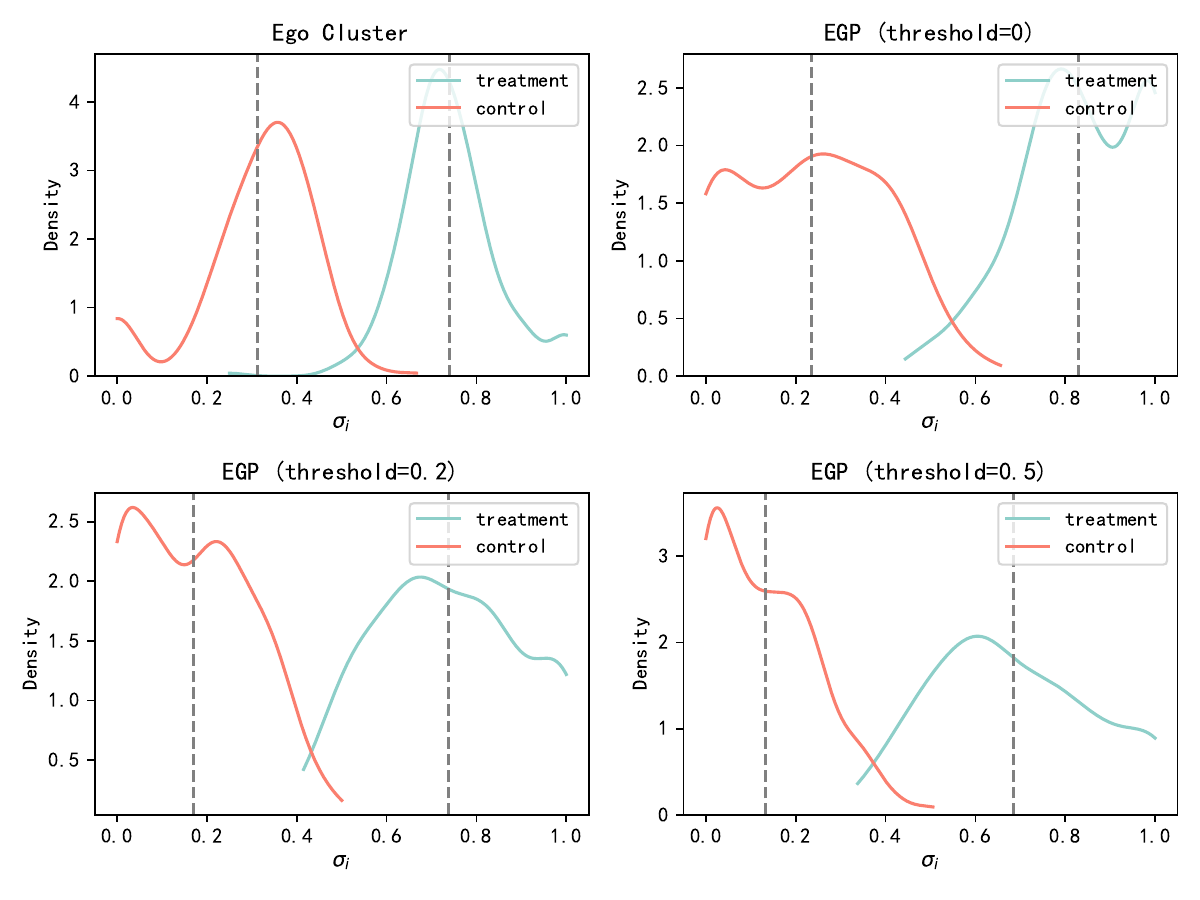}
\caption{The distribution of $\sigma_i$ of different algorithms, which are Ego cluster, Ego partition(threshold=0), Ego partition(threshold=0.2), Ego partition(threshold=0.5).In every subplot, the blue line describes the $\sigma_i's$ distribution of the treatment group, while the yellow line describes the treatment group. And the gray lines perpendicular to the X-axis are the means of the two distributions, respectively.}
\label{sigmafig}
\end{figure*}

Initially, we compare the distribution of $\sigma_i$ for the treatment and control groups across different algorithms, as depicted in Figure \ref{sigmafig}. As can be observed, the disparities in the average treated neighbor ratios $\sigma_i$ between the treatment and control groups for the three variations of the ego group partition are all greater than that of ego cluster, suggesting a reduced bias. Moreover, as the value of $\theta$ escalates, the average treated neighbor ratio $\sigma_i$ within the control group diminishes.

Subsequently, we contrast the bias of the difference-in-means estimator across diverse function types of $g_2$. This includes a linear type, represented by
$$Y_i = 1 + 1 \cdot W_i + 1 \cdot \sigma_i + e_i,$$
and a convex type, represented by
$$Y_i = 2 + 1 \cdot W_i - exp(-3\sigma_i) + e_i.$$
$e_i \sim \mathcal{N}(0,1)$. For each ego approach under consideration, we conduct a Monte Carlo simulation with 1000 iterations to compute the sample mean of the estimators. The average bias of the estimator for different ego approaches, under both linear and convex models, is presented in Table \ref{tab1}.

\begin{table*}[htbp]
\caption{The average bias of the estimator for different ego approaches under both linear and convex models.}\label{tab1}
\begin{center}
\setlength{\tabcolsep}{6mm}{
\begin{tabular}{c|c|c}
  \toprule
  \textbf{Model} & \textbf{Method} & \textbf{Bias}  \\
  \hline
  \multirow{4}*{Linear} & Ego Cluster & -0.565  \\
  & EGP(threshold=0) & -0.416  \\
  & EGP(threshold=0.2) & -0.416  \\
  & EGP(threshold=0.5) & -0.430 \\
  \hline
  \hline
  \multirow{4}*{Convex} & Ego Cluster & -0.614  \\
  & EGP(threshold=0) & -0.463  \\
  & EGP(threshold=0.2) & -0.417  \\
  & EGP(threshold=0.5) & -0.380 \\
  \bottomrule
\end{tabular}
}
\end{center}
\end{table*}

In general, the outcomes align with our initial expectations. The bias of our proposed ego group partition approaches, across different thresholds, is consistently lower than the bias of the ego cluster approach proposed by LinkedIn, under both models. In the context of the linear model, the ego group partition approach with a threshold $\theta=0$ outperforms the other thresholds, as demonstrated in Theorem \ref{thm: r_opt}. For the convex model, the bias of the ego group partition approaches diminishes as the threshold $\theta$ escalates, corroborating our previous hypothesis. In practical scenarios, practitioners can select an appropriate $\theta$ based on the real-world data to optimize the performance of the ego group partition.


\subsection{Real Data}

\begin{table*}[htbp]
\caption{Estimates with re-scaled confidence interval of the metrics in ego group partition experiment and Bernoulli randomized experiment.}\label{tab:real_data}
\begin{center}
\setlength{\tabcolsep}{5mm}{
\begin{tabular}{l|c|c}
  \toprule
   &  ego group partition experiment & Bernoulli randomized experiment  \\
  \hline
  Metric 1 & -0.960\% $\pm$ 0.5\% & -0.499\% $\pm$ 0.602\%   \\
  Metric 2 & -0.488\% $\pm$ 0.5\% & -0.406\% $\pm$ 0.596\%  \\
  \bottomrule
\end{tabular}
}
\end{center}
\end{table*}

We conducted an empirical experiment on Weixin's platform, using ego group partition to estimate the GATE of the algorithm. Weixin is a messaging and calling app that allows users to communicate with friends. One of its built-in features is "Subscriptions", where authors can publish articles and users can read and subscribe to their preferred authors. Upon entering the "Subscriptions" page, users first see articles from their subscribed authors, followed by system-recommended articles from unsubscribed authors. Some of these recommended articles are ones that friends have seen and liked, a feature we refer to as social recommendation. The recommendation team sought to understand if social recommendation incurs social network interference and, if so, the extent of this interference. We conducted a two-week ego group partition experiment and a simple Bernoulli randomized experiment to test for interference. The ego group partition experiment included 96\% of the users, while the Bernoulli randomized experiment included the remaining 4\%. In the treatment group, all socially recommended articles were blocked, while the control group remained unchanged. Both the ego experiment and the Bernoulli randomized experiment had identical treatment and control groups. The underlying hypothesis was that blocking socially recommended articles would decrease both the user's and their social neighbors' activity and consumption on the "Subscriptions" page. We implemented Algorithm 2, i.e., ego group partition with a threshold of $\theta=1$, resulting in an average $\sigma_i$ of 0.6 in the ego treatment group and an average $\sigma_i$ of 0.1 in the ego control group. All data and attributes are collected after user approvals and data masking to protect user privacy.

Table \ref{tab:real_data} provides a detailed breakdown of the outcomes for two key metrics. We re-scaled each metric so that the estimator $\widehat{\tau}_{\text{ego}}$ in the ego group partition experiment would have a confidence interval (CI) width of one. The results corroborate our initial hypotheses. The estimate $\widehat{\tau}_{\text{ego}}$ in the ego group partition experiment demonstrates a more significant decrease compared to the estimate in the Bernoulli randomized experiment. This suggests that social recommendation can indeed cause substantial social network interference.

\section{Discussion}\label{sec5}
The ego group partition algorithms presented in this work offer the potential to estimate the GATE without the need for stringent model assumptions. Compared to graph cluster randomized experiments, our algorithms demonstrate a reduced bias. In contrast to ego cluster experiments, our algorithms generate a greater number of egos, thereby increasing statistical power and addressing the issue of inhomogeneity. Furthermore, these algorithms are well-suited for implementation using parallel computation.

There are several potential avenues for extending our current work. Firstly, the optimal selection of the parameter $\theta$ remains unclear. In practical applications, practitioners typically select a suitable $\theta$ based on historical data and experience. A potential direction for future research could involve expressing the bias of ego group experiments as a function of $\theta$ under specific model assumptions. Optimization techniques could then be employed to provide an optimal solution.

Secondly, our work assumes that the exposure mapping function $f(W_i, \sigma_i)$ is a monotone and convex function with respect to $\sigma_i$. While monotonicity is a common assumption, convexity is not always guaranteed. In certain scenarios, the function $f(W_i, \sigma_i)$ may initially exhibit convexity and then transition to concavity with respect to $\sigma_i$. Further research is warranted to explore whether it's feasible to incorporate additional parameters into the optimization problem \eqref{equa:max1} and achieve the optimal solution under some supplementary assumptions.

Thirdly, the neighborhood interference model, as discussed in this paper, yields a straightforward potential outcome model when applied to real data. To extend our methodology to the k-th order neighborhood interference model, it is necessary to define a property. This property should illustrate the diminishing spillover effects as the distance to the neighbor increases. Understanding this property is crucial as it aids in calculating the bias inherent in the ego group experiment. Furthermore, it provides valuable insights that can guide the optimization of this bias.

Fourthly, the estimator we utilize is the conventional difference-in-means estimator, which solely relies on ego data. There exists a possibility to integrate alter data into the estimator, thereby augmenting the statistical power and identifying consistent estimators for the GATE within the overarching model. A promising avenue for future research could involve examining the asymptotic efficiency of these novel estimators and ascertaining the rate at which variance converges.

These potential enhancements could broaden the applicability of our ego group partition algorithms to more comprehensive models, while also offering corresponding theoretical justifications. We earmark these research trajectories for exploration in future studies.
\\

\noindent\textbf{Code Availability Statement:}
\href{https://github.com/adamdenglu/Ego-Group-Partition}{https://github.com/adamdenglu/Ego-Group-Partition}


\bibliographystyle{ACM-Reference-Format}
\bibliography{ref}


\appendix

\section{Proofs}
\subsection{Proof of Proposition \ref{prop: basemodelbias}}
\begin{proof}
Given the potential outcome model \eqref{equa:linearmodel}, the average outcome of egos in the control group is
\begin{align*}
    \widehat{Y}_C = & \frac{1}{n_0} \sum_{i=1}^n E_i (1-W_i) (\beta_0 + \beta_1 W_i + \beta_2 \sigma_i) \\
    = & \beta_0 + \frac{\beta_2}{n_0} \sum_{i=1}^n E_i (1-W_i) \sigma_i.
\end{align*}
Similarly, the average outcome of egos in the treatment group is
\begin{equation*}
    \widehat{Y}_T = \beta_0 + \beta_1 + \frac{\beta_2}{n_1} \sum_{i=1}^n E_i W_i \sigma_i.
\end{equation*}
So the difference in the average outcome of egos in treatment and control groups is
\begin{equation*}
    \widehat{\tau}_{\text{ego}} =  \widehat{Y}_T - \widehat{Y}_C = \beta_1 + \frac{\beta_2}{n_1} \sum_{i=1}^n E_i W_i \sigma_i - \frac{\beta_2}{n_0} \sum_{i=1}^n E_i (1-W_i) \sigma_i .
\end{equation*}
Consequently, the bias is
\begin{equation*}
   \mathbb{E}(\widehat{\tau}_{\text{ego}}) - \tau = \beta_2 \bigg( \mathbb{E}
   \Big(\frac{1}{n_1} \sum_{i=1}^n E_i W_i \sigma_i - \frac{1}{n_0} \sum_{i=1}^n E_i (1-W_i) \sigma_i\Big) - 1 \bigg).
\end{equation*}

\end{proof}

\subsection{Proof of Proposition \ref{prop: r_ineuqa}}\label{sub:prof_r_ineuqa}
\begin{proof}
Recall that $R = \frac{1}{n_1} \sum_{i=1}^n E_i W_i \sigma_i - \frac{1}{n_0} \sum_{i=1}^n E_i (1-W_i) \sigma_i$, where $\sigma_i =  \sum_{j=1}^n \frac{A_{ij}}{d_i} W_j$. For any $i$, $0 \leq \sigma_i \leq 1$. Besides, $n_1$ and $n_0$ are the numbers of egos in treatment and control groups, so
\begin{equation*}
    n_1 = \frac{1}{n_1} \sum_{i=1}^n E_i W_i, \quad n_0 = \frac{1}{n_0} \sum_{i=1}^n E_i (1-W_i).
\end{equation*}
Hence, we can deduce that
\begin{equation*}
    R \leq \frac{1}{n_1} \sum_{i=1}^n E_i W_i = 1,
\end{equation*}
and 
\begin{equation*}
    R \geq - \frac{1}{n_0} \sum_{i=1}^n E_i (1-W_i) = -1.
\end{equation*}
Combine these two inequation, we have
\begin{equation*}
    -1 \leq R \leq 1.
\end{equation*}
\end{proof}

\subsection{Proof of Theorem \ref{thm: r_opt}}
\begin{proof}
The optimization problem is
\begin{equation*}
    \max\limits_{ \{W_j : E_j = 0\} } \mathbb{E}\big[R \,\,|\,\, \{W_i: E_i=1 \}\big].
\end{equation*}

Note that we can split $\sigma_i$ into two parts, one contributed by ego friends and the other by non-ego friends:
\begin{equation*}
\sigma_i =  \sum_{j=1}^n \frac{A_{ij}}{d_i} W_j  =  \sum_{j=1}^n \frac{A_{ij}}{d_i} E_j W_j + \sum_{j=1}^n \frac{A_{ij}}{d_i} (1-E_j) W_j .
\end{equation*}
Then
\begin{align*}
      & \frac{1}{n_1} \sum_{i=1}^n E_i W_i \sigma_i \\
    = & \,\, \frac{1}{n_1} \sum_{i=1}^n E_i W_i \sum_{j=1}^n \frac{A_{ij}}{d_i} W_j  \\
    = & \,\, \frac{1}{n_1} \sum_{i=1}^n E_i W_i \bigg( \sum_{j=1}^n \frac{A_{ij}}{d_i} E_j W_j + \sum_{j=1}^n \frac{A_{ij}}{d_i} (1-E_j) W_j \bigg), \\
\end{align*}
and 
\begin{align*}
      & \frac{1}{n_0} \sum_{i=1}^n E_i (1-W_i) \sigma_i \\
    = & \,\, \frac{1}{n_0} \sum_{i=1}^n E_i (1-W_i) \bigg( \sum_{j=1}^n \frac{A_{ij}}{d_i} E_j W_j + \sum_{j=1}^n \frac{A_{ij}}{d_i} (1-E_j) W_j \bigg). \\
\end{align*}
Therefore, 
\begin{equation*}
    R =  \frac{1}{n_1} \sum_{i=1}^n E_i W_i \sigma_i - \frac{1}{n_0} \sum_{i=1}^n E_i (1-W_i) \sigma_i \overset{\Delta}{=} \Lambda_1 + \Lambda_2,
\end{equation*}
where
\begin{equation*}
\Lambda_1 =  \frac{1}{n_1} \sum_{i=1}^n \sum_{j=1}^n  E_i W_i \frac{A_{ij}}{d_i} E_j W_j - \frac{1}{n_0} \sum_{i=1}^n \sum_{j=1}^n  E_i (1-W_i) \frac{A_{ij}}{d_i} E_j W_j ,
\end{equation*}
and
\begin{equation*}
\begin{aligned} 
    \Lambda_2 = & \,\, \frac{1}{n_1} \sum_{i=1}^n \sum_{j=1}^n  E_i W_i \frac{A_{ij}}{d_i} (1-E_j) W_j  \\
    & \,\, - \frac{1}{n_0} \sum_{i=1}^n \sum_{j=1}^n  E_i (1-W_i) \frac{A_{ij}}{d_i} (1-E_j) W_j.
\end{aligned}
\end{equation*}

Note that $\mathbb{E}[R | \{W_i: E_i=1 \}] = \mathbb{E}[\Lambda_1 | \{W_i: E_i=1 \}] + \mathbb{E}[\Lambda_2 | \{W_i: E_i=1 \}] $. In $\Lambda_1$ all $W_i$s are egos' treatment assignments, which are fixed when we allocate egos' neighbors' treatment. So $ \mathbb{E}[\Lambda_1 | \{W_i: E_i=1 \}] = \Lambda_1$, and we only need to maximize $\mathbb{E}[\Lambda_2 | \{W_i: E_i=1 \}]$. 
Let's rewrite $\Lambda_2$ as the summation of $(1-E_j)W_j$:
\begin{align*}\label{equa:lambda2}
    \Lambda_2 = & \,\, \sum_{j=1}^n (1-E_j) W_j \sum_{i=1}^n  \frac{ E_i W_i A_{ij} }{ n_1 d_i } \nonumber \\
    & \,\, -  \sum_{j=1}^n (1-E_j) W_j \sum_{i=1}^n \frac{ E_i (1-W_i) A_{ij} }{n_0 d_i } \nonumber \\
    =  & \,\, \sum_{j=1}^n (1-E_j) W_j \bigg( \sum_{i=1}^n \frac{ E_i W_i A_{ij} }{ n_1 d_i } - \sum_{i=1}^n  \frac{ E_i (1-W_i) A_{ij} }{n_0 d_i } \bigg) \nonumber \\
    =  & \,\, \sum_{j=1}^n (1-E_j) W_j \Delta_j 
\end{align*}

Since all $W_j$ in the summation of $\Delta_j$ satisfy $E_j = 1$, $\Delta_j$ is fixed. Hence
\begin{align*}
\mathbb{E}[\Lambda_2 | \{W_i: E_i=1 \}] = \,\, & \mathbb{E}\Big[\sum_{j=1}^n (1-E_j) W_j \Delta_j  | \{W_i: E_i=1 \}\Big]\\
\leq & \,\, \mathbb{E}\Big[\sum_{j=1}^n (1-E_j) \max \{\Delta_j, 0\} | \{W_i: E_i=1 \}\Big]\\
= & \,\, \sum_{j=1}^n (1-E_j) \max \{\Delta_j, 0\}.
\end{align*}
The inequality holds if and only if $W_j = \mathcal{I}\{\Delta_j > 0\}$ for those $j$ with $E_j =0$. In other words, to maximize $\Lambda_2$, we can set $W_j = 1$ when $\Delta_j > 0$; set it to 0 otherwise.

\end{proof}

\subsection{Proof of Theorem \ref{thm: conv_opt}}
\begin{proof}
The optimization problem is
\begin{equation*}
    \max\limits_{ \{W_j : E_j = 0 \}}  \mathbb{E}(\sigma_i|W_i=1) - (1 + \theta) \mathbb{E}(\sigma_i|W_i=0).
\end{equation*}

Denote $\tilde{R} = \frac{1}{n_1} \sum_{i=1}^n E_i W_i \sigma_i - \frac{1+\theta}{n_0} \sum_{i=1}^n E_i (1-W_i) \sigma_i$. Following the same steps in \ref{sub:prof_r_ineuqa}, we can obtain $\tilde{R} = \tilde{\Lambda}_1 + \tilde{\Lambda}_2$,
where 
\begin{equation*}
\tilde{\Lambda}_1 =  \frac{1}{n_1} \sum_{i=1}^n \sum_{j=1}^n  E_i W_i \frac{A_{ij}}{d_i} E_j W_j - \frac{1 + \theta}{n_0} \sum_{i=1}^n \sum_{j=1}^n  E_i (1-W_i) \frac{A_{ij}}{d_i} E_j W_j ,
\end{equation*}
and
\begin{equation*}
\begin{aligned} 
    \tilde{\Lambda}_2 = & \,\, \frac{1}{n_1} \sum_{i=1}^n \sum_{j=1}^n  E_i W_i \frac{A_{ij}}{d_i} (1-E_j) W_j  \\
    & \,\, - \frac{1 + \theta}{n_0} \sum_{i=1}^n \sum_{j=1}^n  E_i (1-W_i) \frac{A_{ij}}{d_i} (1-E_j) W_j.
\end{aligned}
\end{equation*}

Again, we only need to maximize $\mathbb{E}[\tilde{\Lambda}_2 | \{W_i: E_i=1 \}]$. Let's rewrite $\tilde{\Lambda}_2$ as the summation of $(1-E_j)W_j$:
\begin{align*}\label{equa:lambda2}
    \tilde{\Lambda}_2 = & \,\, \sum_{j=1}^n (1-E_j) W_j \sum_{i=1}^n  \frac{ E_i W_i A_{ij} }{ n_1 d_i } \nonumber \\
    & \,\, -  (1+\theta)\sum_{j=1}^n (1-E_j) W_j \sum_{i=1}^n \frac{ E_i (1-W_i) A_{ij} }{n_0 d_i } \nonumber \\
    =  & \,\, \sum_{j=1}^n (1-E_j) W_j \bigg( \sum_{i=1}^n \frac{ E_i W_i A_{ij} }{ n_1 d_i } - (1+\theta) \sum_{i=1}^n  \frac{ E_i (1-W_i) A_{ij} }{n_0 d_i } \bigg) \nonumber \\
    \overset{\Delta}{=}  & \,\, \sum_{j=1}^n (1-E_j) W_j \Delta^{\prime}_j 
\end{align*}

Since all $W_j$ in the summation of $\Delta^{\prime}_j$ satisfy $E_j = 1$, $\Delta^{\prime}_j$ is fixed. Hence
\begin{align*}
\mathbb{E}[\Lambda_2 | \{W_i: E_i=1 \}] = \,\, & \mathbb{E}\Big[\sum_{j=1}^n (1-E_j) W_j \Delta^{\prime}_j  | \{W_i: E_i=1 \}\Big]\\
\leq & \,\, \mathbb{E}\Big[\sum_{j=1}^n (1-E_j) \max \{\Delta^{\prime}_j, 0\} | \{W_i: E_i=1 \}\Big]\\
= & \,\, \sum_{j=1}^n (1-E_j) \max \{\Delta^{\prime}_j, 0\}.
\end{align*}
The inequality holds if and only if $W_j = \mathcal{I}\{\Delta^{\prime}_j > 0\}$ for those $j$ with $E_j =0$. In other words, to maximize $\Lambda_2$, we can set $W_j = 1$ when $\Delta^{\prime}_j > 0$; set it to 0 otherwise.

The last step is to show that $\Delta^{\prime}_j = > 0$ is equivalent to $\tilde{\Delta}_j >0$, which is trivial.
\end{proof}

\end{document}